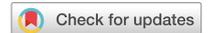

# Highly sensitive multicore fiber accelerometer for low frequency vibration sensing

Josu Amorebieta[1]✉, Angel Ortega-Gomez[1], Gaizka Durana[1], Rubén Fernández[1], Enrique Antonio-Lopez[2], Axel Schülzgen[2], Joseba Zubia[1], Rodrigo Amezcua-Correa[2] & Joel Villatoro[1,3]✉

We report on a compact, highly sensitive all-fiber accelerometer suitable for low frequency and low amplitude vibration sensing. The sensing elements in the device are two short segments of strongly coupled asymmetric multicore fiber (MCF) fusion spliced at 180º with respect to each other. Such segments of MCF are sandwiched between standard single mode fibers. The reflection spectrum of the device exhibits a narrow spectrum whose height and position in wavelength changes when it is subjected to vibrations. The interrogation of the accelerometer was carried out by a spectrometer and a photodetector to measure simultaneously wavelength shift and light power variations. The device was subjected to a wide range of vibration frequencies, from 1 mHz to 30 Hz, and accelerations from 0.76 mg to 29.64 mg, and performed linearly, with a sensitivity of 2.213 nW/mg. Therefore, we believe the accelerometer reported here may represent an alternative to existing electronic and optical accelerometers, especially for low frequency and amplitude vibrations, thanks to its compactness, simplicity, cost-effectiveness, implementation easiness and high sensitivity.

For a long time, accelerometers have been used to detect and measure vibrations with high sensitivity and precision. Thus, they have a wide variety of applications. For instance, in the heavy industry, accelerometers are used to monitor low-frequency vibrations in large rotating machineries or in oil pipes[1], or in structural health monitoring, to supervise the condition of pillars, bridges, etc.[2]. They are also used in biomedicine and biomechanics[3], and even in gravitational wave detectors[4]. Accelerometers are one of the key elements in seismology as well[5], where they are used for the detection and monitoring of ground motions caused by earthquakes, volcanic eruptions, explosions, landslides, tsunamis, avalanches, etc. In these cases, accelerometers with high sensitivity for low amplitude vibrations are required.

The detection of vibrations with low frequencies is a very challenging field. For example, the frequency range of ground motions caused by natural events or explosions is between 0.1 and 20 Hz[6,7]. For the case of tsunamis, such frequency range is even narrower, from 0.1 to 1 Hz[8]. Moreover, it is important to identify the ground motions accurately from the surrounding noise. Therefore, accelerometers for such applications must be highly sensitive and must be capable of measuring acceleration in a wide range. Additional requirements for accelerometers include simple operation, compactness, robustness, capability to operate in hostile or harsh environments and multi-point sensing[9,10]. Finally, as natural events are usually unpredictable and spaced in time[11], such accelerometers must be reliable, long-lasting, and should require minimum or no maintenance.

So far, the most spread accelerometers for low frequencies are based on piezoelectric components[12], MEMS membranes[13] or pendulums[14] that move with vibrations, or have electrochemical nature[15]. Moreover, frequently, such operating principles are combined to enhance their overall performance[7]. The technology of electronic accelerometers is very mature and cost-effective. However, the harsh environments in which these accelerometers are commonly deployed, such as seabed or boreholes, may affect the lifetime of their elements. For such applications, accelerometers based on optical fibers are a good alternative. Fiber-based accelerometers have important advantages that include small size, electromagnetic immunity, as they do not require any electric component to operate, high resolution, remote and long-distance operation capabilities.

[1]Department of Communications Engineering, University of the Basque Country UPV/EHU, 48013 Bilbao, Spain. [2]CREOL -The College of Optics and Photonics, University of Central Florida, Orlando, FL 162700, USA. [3]Ikerbasque-Basque Foundation for Science, 48011 Bilbao, Spain. ✉email: josu.amorebieta@ehu.eus; agustinjoel.villatoro@ehu.es

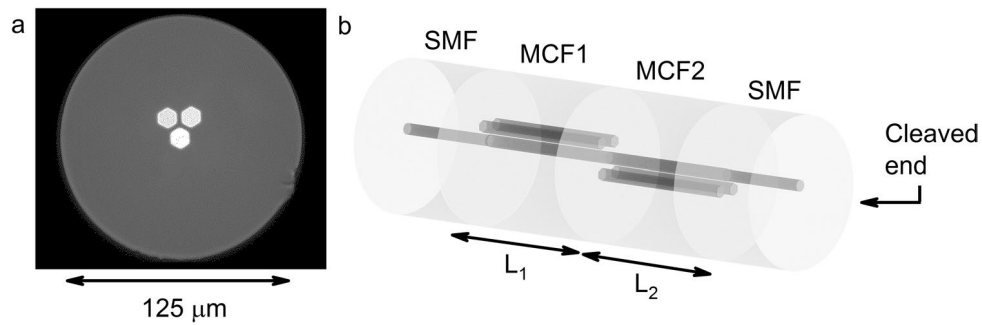

**Figure 1.** (**a**) Cross section of the asymmetric MCF. (**b**) Schematic layout of the device drawn with Blender v2.82 (https://www.blender.org/).

Among optical fiber accelerometers, those based on interferometry and fiber Bragg gratings (FBGs) are the most advanced configurations. Optical fiber interferometric accelerometers feature larger dynamic range, wider frequency response band and higher sensitivity compared to some electronic accelerometers[16–18]. In fact, optical fiber interferometric seismometers capable of detecting vibrations of few mHz have been reported[19,20]. However, they are bulky and their interrogation tends to be complex. FBG-based optical accelerometers are more compact. Moreover, they provide high sensitivity, large dynamic range and multiplexing ability[21]; and may operate in frequency response bands below 1 Hz[22,23]. To reach such performance, they require sophisticated interrogation systems that entail picometer-resolution interrogators. In addition, they require elaborated packaging. As a consequence, FBG accelerometers are expensive.

In recent years, multicore fibers (MCFs) have drawn much attention as multipurpose sensing elements[24]. As accelerometers, strongly coupled MCFs have proved to have much potential[25]. Moreover, their capability to withstand and operate under elevated strain and temperature conditions has been reported as well[26,27], which is a demanded characteristic for harsh environments or outdoors implementations.

In this work, we report on a highly sensitive all-fiber optical accelerometer suitable for sensing vibrations of extremely low frequencies (down to 1 MHz) and low amplitudes. The device is compact and consists of two segments of MCF sandwiched between standard single mode fiber. The MCF segments have different lengths and are rotated 180° with respect to each other. Due to its architecture, the reflection spectrum of the device exhibits a narrow peak that shrinks when it is subjected to vibrations. To test the device, it was subjected to vibrations from 1 mHz to 30 Hz and accelerations from 0.76 to 29.64 mg. The performance of our device was compared and calibrated with a commercial electronic accelerometer. We believe that the simplicity and high performance of the MCF accelerometer reported here are appealing for several applications; particularly those where frequencies are low.

## Sensor design, operating mechanism and fabrication

The MCF used to build the accelerometer was fabricated at the University of Central Florida (Orlando, USA). It is an asymmetric strongly coupled MCF consisting of three cores, where one of the cores is located at the geometrical center of the fiber, whereas the other two are surrounding it and arranged adjacently in a V-like configuration (Fig. 1a). Each core is made of Germanium doped silica, and has a mean diameter of 9 μm and a numerical aperture (NA) of 0.14 at 1550 nm to match with that of the SMF. The cores are separated 11.5 μm from each other and embedded in a pure silica cladding of 125 μm of diameter.

The architecture of the device is sketched in Fig. 1b. The sensor consists of two cascaded short segments of different lengths of MCF rotated 180° with respect to each other and sandwiched between two SMFs, resulting in a SMF-MCF1-MCF2-SMF structure. In this structure, the distal SMF has a cleaved end that acts as a low reflectivity mirror in order the device to be interrogated in reflection mode. The benefits of such structure will be discussed throughout this section.

The theoretical background of strongly coupled MCFs relies on the coupled mode theory (CMT)[28–31]. According to it, if at least two waveguides are close enough to interact, a cyclical power transfer between the waveguides will take place due to the overlapping between the propagating modes through each of them. For conventional CMT, it is assumed that the propagating modes under study are orthogonal[32]. In the simplest case, if we assume two single mode waveguides named 1 and 2 that are so close to each other that the evanescent field from one guide penetrates into the other, there is a coupling between the two propagating modes. For waveguide 1, such propagation can be expressed as:

$$\frac{\partial a_1}{\partial z} = -j(\beta_1 + k_{1,1})a_1 - jk_{1,2}a_2 \qquad (1)$$

where $a$ is the amplitude of the mode in the waveguide indicated in the subindex, $\beta$ is its corresponding propagation constant and the $k$ parameters are the mutual and self-coupling coefficients between the orthogonal propagating modes in the waveguides 1 and 2, respectively, along the $z$ axis where the propagation is taking place. Identical expression is valid for the propagation in waveguide 2 by substituting in Eq. (1) the subindex 1 for 2 and vice versa.

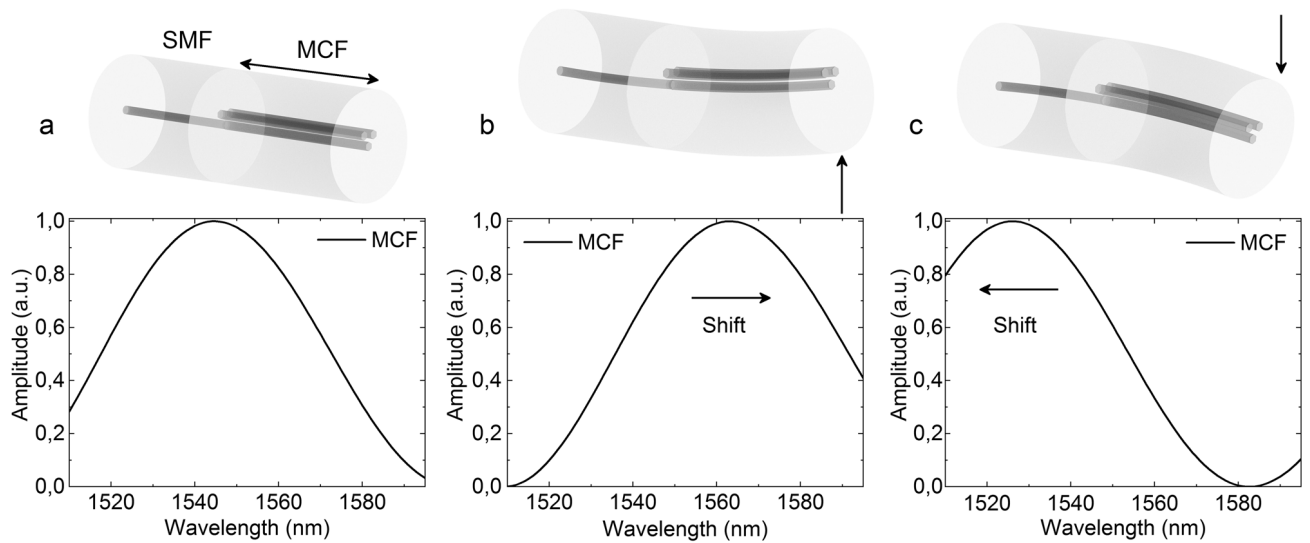

**Figure 2.** Simulated spectra for the cases in which a segment of MCF is (**a**) straight, (**b**) bent upwards and (**c**) bent downwards when the cores are positioned in a V-like configuration. The arrow indicates the bending direction and the wavelength shift in each case.

Now let us assume the boundary condition in which the amplitude of the mode $a$ only exists in one of the waveguides at $z=0$. Thus, by applying the condition $a_1(0)=1$ and $a_2(0)=0$, it is possible to calculate the coupled power in any of the waveguides at any distance by calculating $P(z) = a(z) * a^*(z)$, where $a^*$ refers to the conjugate amplitude of the mode. For such case, the normalized coupled power in the waveguide 1 at a certain propagation distance $z$, can be expressed as:

$$P_1(z) = cos^2(Sz) + cos^2(\gamma)sin^2(Sz) \quad (2)$$

where $S = \sqrt{\delta^2 + k^2}$, $\delta = (\beta_1 - \beta_2)/2$ and $\tan(\gamma) = k/\delta$.

In strongly coupled MCFs, each of the cores acts as a waveguide. In such coupled structures, the propagating modes are called supermodes[33,34], which are the linear combination of the propagating modes through each of the individual waveguides. When such MCFs are excited in their central core by the incoming $LP_{01}$ mode from the SMF, the two orthogonal supermodes that have power in the central core are coupled. Such supermodes are named $SP_{01}$ and $SP_{02}$, and are specific for each MCF. Moreover, for strongly coupled MCFs as the one employed to manufacture this accelerometer, in which all the cores are identical in terms of size and physical properties, and the distance between the central and the neighboring cores remains unaltered, this supermode coupling provokes the power distribution among all the adjacent cores to be identical. Therefore, particularizing Eq. (2) for a stub of the MCF in Fig. 1a, the normalized coupled power in the central core can be expressed as:

$$P(z) = cos^2\left(\frac{\sqrt{3}\pi \Delta n}{\lambda}z\right) + \frac{1}{3}sin^2\left(\frac{\sqrt{3}\pi \Delta n}{\lambda}z\right) \quad (3)$$

where $\Delta n$ is the difference between the effective refractive indexes of the two propagating coupled supermodes and depends on the physical characteristics of the MCF, $\lambda$ is the excitation wavelength, and $z$ is the distance at which the normalized power is being evaluated along the propagation axis. Therefore, the transmitted power will vary periodically, with a maximum at certain values of $z$ and a minimum at others.

Now, if the length of MCF is fixed, let us say L, the transmission of an SMF-MCF-SMF structure can also be described by particularizing Eq. (3) for $z=L$. If such a structure is excited with a broadband source, the transmission spectrum will be periodic in wavelength according to the phase in Eq. (3).

When an MCF is bent, each core suffers different levels of strain, and their respective refractive indices vary accordingly[35–39], modifying the effective refractive indices of the two propagating supermodes and therefore, the power coupling conditions, which will be reflected in the spectrum. In our case, this effect, added to the asymmetrical arrangement of the cores and their orientation, will cause detectable wavelength shift and coupled power variations that will have unique characteristics depending on the applied bending direction and amplitude, making the MCF ideal for direction-sensitive bending sensors. As demonstrated in[40], when the position of the cores and the applied bending are aligned as in Fig. 2, where the cores are orientated in a V-like configuration and the MCF is bent upwards and downwards, only wavelength shifts will be noticed in the spectrum; whereas if we rotate the fiber 90°, only coupled power variations will be noticed.

As a step forward of such operating principle, the accelerometer proposed in this work consists of two short segments of the aforementioned MCF (MCF1 and MCF2) of similar but different lengths ($L_1$ and $L_2$) that are cascaded and rotated 180° with respect to each other. For this configuration, Eq. (1) has to be applied to each segment. Hence, the normalized output power of the cascade is the product of the individual normalized power outputs of each MCF segment:

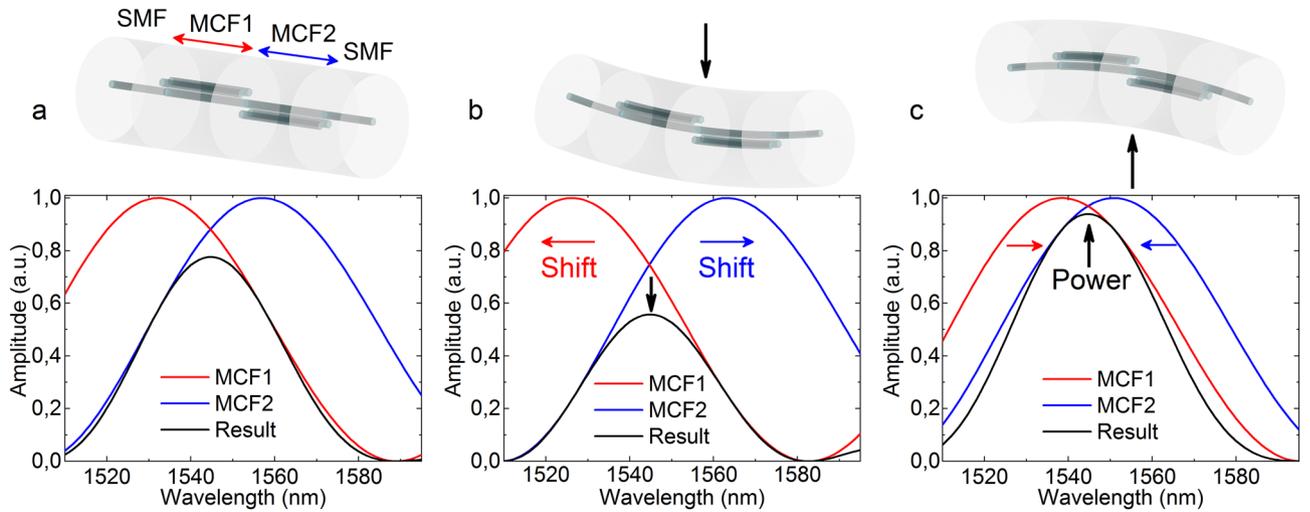

**Figure 3.** Simulated spectra of each of the MCF segments and the resulting spectra for the cases where the structure is (**a**) straight, (**b**) bent upwards and (**c**) bent downwards by its fusion splice point. The arrow indicates the bending direction, the wavelength shift or the power variation in each case. The cores of MCF1 are in a V-like configuration, whereas the ones in MCF2 are in an inverted V-like configuration.

$$P(L_1, L_2) = P_{MCF1}(L_1) * P_{MCF2}(L_2) \quad (4)$$

where the subindexes MCF1 and MCF2 are referred to each short MCF segment of lengths $L_1$ and $L_2$, respectively. Thus, for $\theta = \sqrt{3}\pi \Delta n / \lambda$, the normalized output power in the central core after passing through the two MCF segments is as follows:

$$P(L_1, L_2) = cos^2(\theta L_1) * cos^2(\theta L_2) + \frac{1}{9}sin^2(\theta L_1) * sin^2(\theta L_2) + \frac{1}{3}cos^2(\theta L_1) * sin^2(\theta L_2) + \frac{1}{3}cos^2(\theta L_2) * sin^2(\theta L_1) \quad (5)$$

If we compare the predominant terms in Eqs. (3) and (5), in Eq. (3) it is a squared cosine whereas in Eq. (5) it can be considered a cosine raised to the fourth. Thus, a spectrum derived from Eq. (5) will have narrower peak or peaks than one from Eq. (3) for identical MCF lengths. Moreover, the visibility of a spectrum from Eq. (5) will be higher as well, as the contribution of the rest of the terms in the equation is less than the contribution of the term in Eq. (3), which makes the difference between adjacent maxima and minima to be lower in the latter. Hence, the advantages of cascading two MCF segments compared to a single MCF segment are narrower peaks in the spectrum and higher visibility, which facilitate tracking any change in it.

By operating in reflection mode, the normalized output power is the product of Eq. (5) by itself due to the back-and-forth path of the light through the SMF-MCF1-MCF2-SMF structure; so it can be assumed that the predominant term is a cosine raised to the eighth. Thus, this is an easy manner to improve the narrowness and visibility of the spectrum even more and the reason why this device operates in such configuration.

Regarding the fiber arrangement, by rotating the two MCF segments 180° with respect to each other, each of them will show contrary behavior in terms of wavelength shift and amplitude of the spectrum when they are bent due to their direction sensitive nature that has been explained previously. When the position of the cores of each MCF segment and the applied vertical bending are aligned as in Fig. 3, where one of the MCF segments has its cores oriented in a V-like configuration and the other MCF segment has its cores oriented in an inverted V-like configuration (or rotated 180°), only pronounced amplitude variations will take place in the spectrum. In order the device to perform as shown in Fig. 3, MCF segments of different lengths are compulsory to avoid any ambiguity in the measurement. If the lengths were identical, the spectra of both segments would be overlapped in idle state, being that situation the point at which the maximum reflected light power would take place. Each spectrum would shift in opposite directions when the structure was bent, but only power decreases would be recorded, resulting in the same or similar power readings for opposite bending directions. Such ambiguity or loss in sensitivity is avoided by using segments of different lengths, as for this case, the measured power increases and decreases accordingly with the applied bending direction compared to the power measurement in idle state. Such amplitude variations in the spectrum are proportional to power variations, and therefore, only a PD will be necessary to interrogate the device. Such simplicity makes this SMF-MCF1-MCF2-SMF structure appealing as a very sensitive and cost-effective accelerometer, as it does not require high performance or ad-hoc equipment to operate.

To manufacture a device with such characteristics, some design constraints were required to be considered: Its spectrum had to be confined within the interrogation window (from 1510 to 1595 nm, according to our interrogation setup) at any time and it must have a unique and well-defined peak with no secondary lobes. Such requirements are mandatory to minimize any sensitivity loss when measuring the reflected light power that is caused by adjacent lobes with opposite trends (one increases whereas the other decreases) in the same interrogation window, as shown in[40].

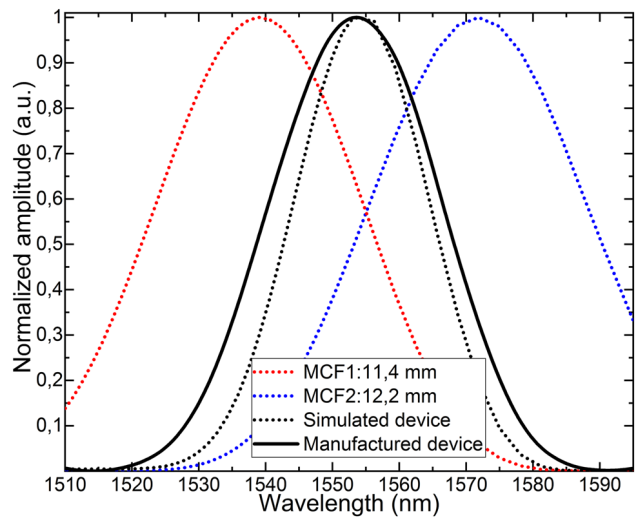

**Figure 4.** Normalized spectra of the simulated (black dashed line) and manufactured devices (black continuous line). Notice that the maxima of both curves is around 1554 nm and there are no secondary lobes. Simulated spectra of MCF segments of 11.4 mm (red dashed line) and 12.2 (blue dashed line) are shown as well. As indicated in Eq. (2) their product results in the black dashed line.

The best fitting lengths for the MCF segments that fulfilled the requirements were 11.4 mm and 12.2 mm, resulting in a compact device of 23.6 mm. To manufacture such device (illustrated in Fig. 1b), a precision fiber cleaver (Fujikura CT-105) and a specialty fiber fusion splicer (Fujikura 100P+) were used. On the one hand, the cleaver allowed us to cut MCF segments of the desired length with 10 μm precision. On the other hand, the splicer can align the central core of the MCF with the unique of the SMF with high precision and has a rotating mechanism and an imaging system that allows observing the end-face of the MCF. Once the MCF segments were rotated 180° with respect to each other, they were spliced, so the central cores of all the segments of the structure were aligned. The fabrication process of our device is inexpensive, fast, and reproducible.

The spectra of the simulated and the manufactured devices are shown in Fig. 4, along with the simulation for each of the MCF segments that comprise the structure. Such simulations were carried out with PhotonDesign simulation software. In such figure, it can be noticed that the curves corresponding to the manufactured and simulated devices agree well, and that the design constraints that included one well-defined and centered peak with no secondary lobes were achieved.

## Results and discussion

The interrogation of the device is simple and was carried out with commercial equipment. It consists of a broadband light source (Safibra, s.r.o.) centered at 1550 nm and an InGaAs PD (Thorlabs PDA30B2). To interrogate the device in reflection mode, a fiber optic circulator was used. As it can be noticed in the simulations in Fig. 3, when the structure is bent by the point in which both MCFs are fusion spliced to each other and with that specific core orientation, only power variation will take place. However, when the physical device is subjected to the same effect, a slight wavelength shift is likely to happen as well apart from the amplitude variation. This is caused by two factors: in first place, the impossibility to apply the bending only and exactly at the fusion splice point; and in second place, the length difference of 0.8 mm between the MCF segments, which will cause a small variation in the shift of each against the same stimulus. Due to that, the device was also interrogated with a spectrometer (Ibsen Photonics I-MON-512 High Speed) to monitor the wavelength shift in the spectrum. Such measurement was used as an indicator of the relation between the direction of the applied bending and the position of the cores, as according to Fig. 3, small wavelength shifts would imply the accelerometer is operating as intended, as it is optimized to maximize the power variation.

To test the device, a horizontally fixed rectangular methacrylate thin plate was used. Underneath and at the center of it, an amplified piezoelectric actuator (Thorlabs APFH720 combined with Thorlabs MDT694B amplifier) was fixed so that the plate could vibrate only in the vertical plane. The piezoelectric actuator was connected to a function generator (Keysight Technologies 33220A) to generate signals of diverse amplitudes and frequencies. Then, the manufactured device was surface bonded with cyanoacrylate adhesive to the upper side of the plate, locating the MCF1-MCF2 splice at the center of it and just above the piezoelectric actuator, as it can be observed in the scheme of the experimental setup shown in Fig. 5. It was surface bonded with its cores oriented as in Fig. 3 to match the direction of vibration. Adjacent to the device, a commercial accelerometer (Pico Technology PP877 with Pico Technology TA096) was fixed for comparison and calibration purposes, as this electronic accelerometer provided the relation between the amplitude of the vibration and the acceleration. All the tests were carried out at room temperature (25 °C) and the raw signal of the time response of both devices was monitored and recorded.

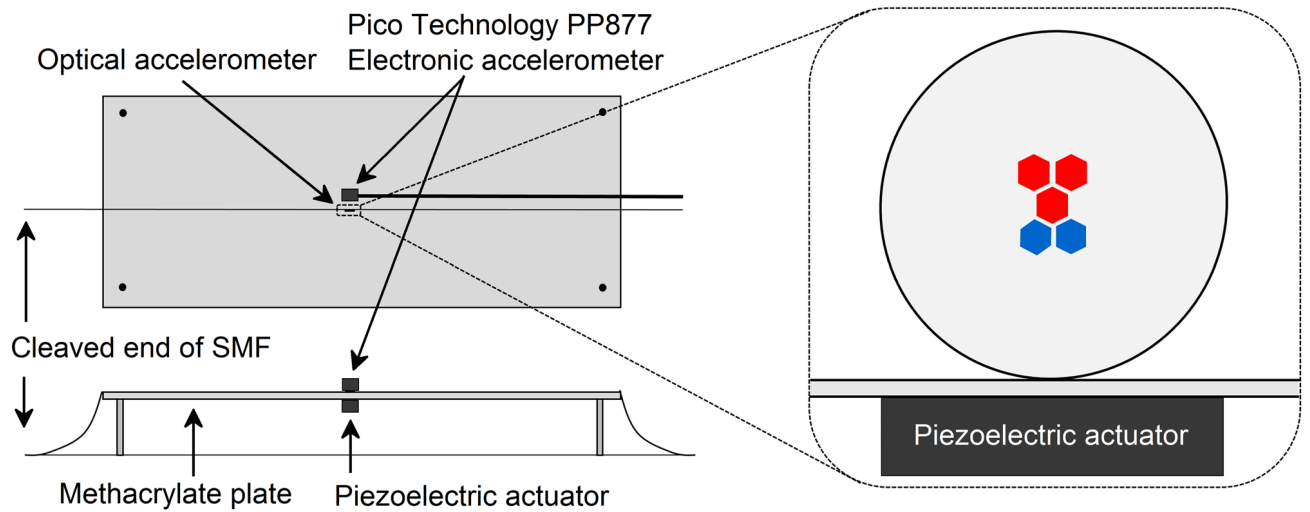

**Figure 5.** Schematic lateral and top views of the experimental setup drawn with Origin2019b (https://www.originlab.com/). The close-up shows how the manufactured optical accelerometer was surface bonded to the plate. Red cores belong to MCF1 whereas blue cores belong to MCF2. The red central core indicates MCF1 is in front of MCF2, as they share common central core. Adjacent to it, the Pico Technology PP877 electronic accelerometer was fixed.

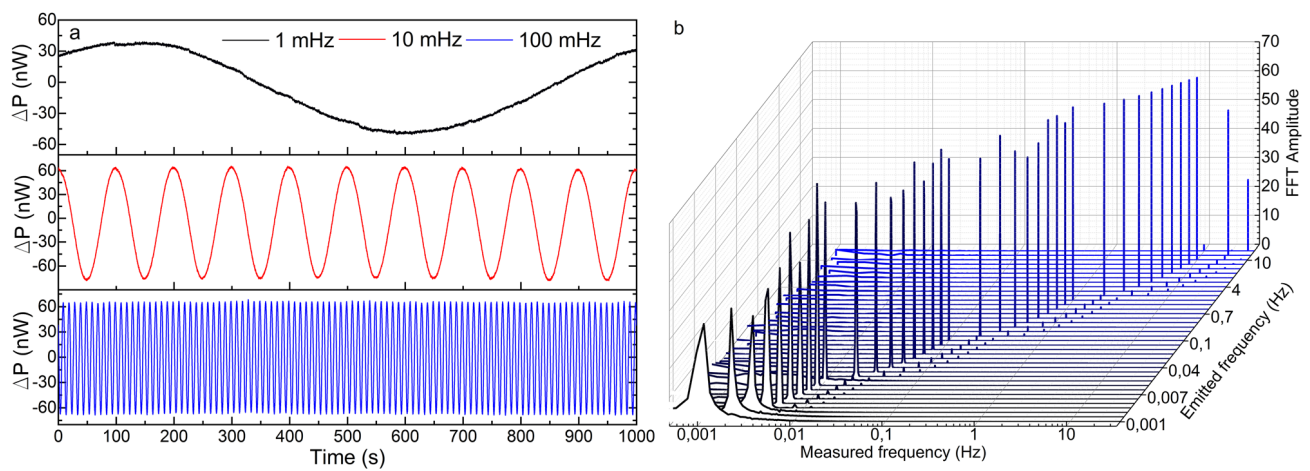

**Figure 6.** Results of the power measurements in the manufactured optical device. (**a**) Time response of three representative cases. (**b**) FFT amplitudes for frequencies from 30 Hz down to 1 mHz for a sinusoidal signal of 1 Vpp. The Measured frequency axis is in logarithmic scale.

According to the optical accelerometer, the time response signals in terms of wavelength at which the maxima in the spectrum takes place ($\lambda$) and measured power in the PD (P) were acquired. The value of such parameters with the device in idle state were taken as reference ($\lambda_{ref}$, $P_{ref}$) to obtain the wavelength shift ($\Delta\lambda = \lambda - \lambda_{ref}$) and power variation ($\Delta P = P - P_{ref}$), respectively. Subsequently, the FFT of such signals was done to obtain the amplitude of their corresponding frequency components and weights. The criteria to define the limit of detection (LoD) was set to be a signal to noise ratio (SNR) of 3 in the FFT amplitude of the most prominent component, which is commonly taken as a rule[41].

The first test consisted in emitting a sinusoidal signal of 1 Vpp amplitude and varying its frequency from 30 Hz down to 1 MHz (the lowest frequency provided by the function generator) in several steps so that the LoD in terms of frequency of each device could be defined. The results are shown from Figs. 6 to 8. The manufactured device detected every vibration clearly down to 1 mHz in wavelength shift and power variation (see Figs. 6a and 7a). The small wavelength shift in Fig. 7a indicates that the device has been surface bonded with the proper core orientation to the plate, and explains the fact that the FFT amplitudes are lower for the wavelength shift measurements than those for the power variation. Nevertheless, even in this configuration aimed at maximizing the power variation, the device has detected such low vibrations by its wavelength shift as well, which is an indicator of its high sensitivity.

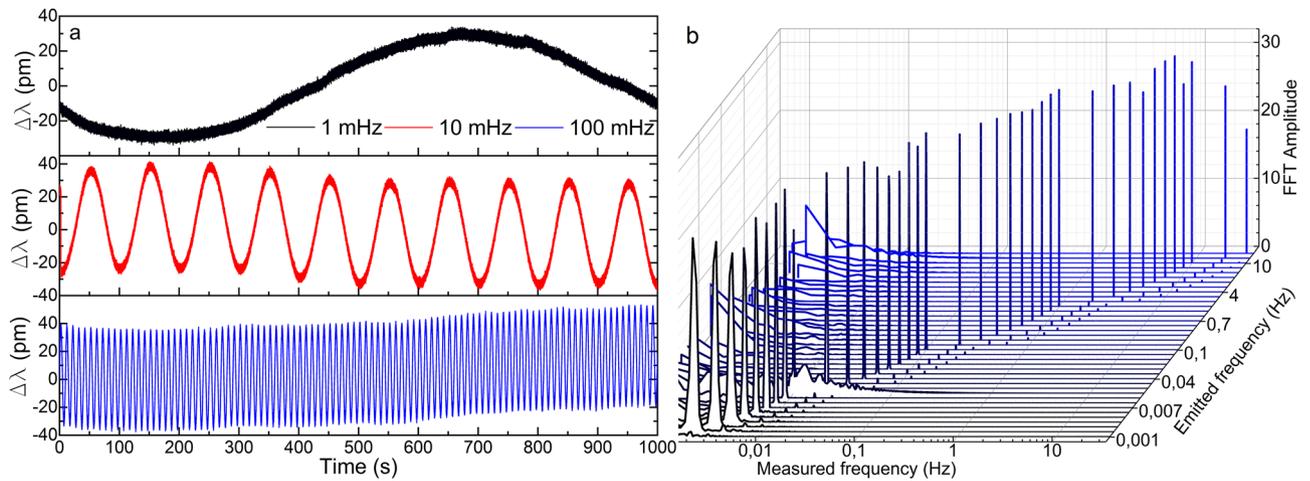

**Figure 7.** Results of the wavelength shift measurements in the manufactured optical device. (**a**) Time response of three representative cases. (**b**) FFT amplitudes for frequencies from 30 Hz down to 1 mHz for a sinusoidal signal of 1 Vpp. The Measured frequency axis is in logarithmic scale.

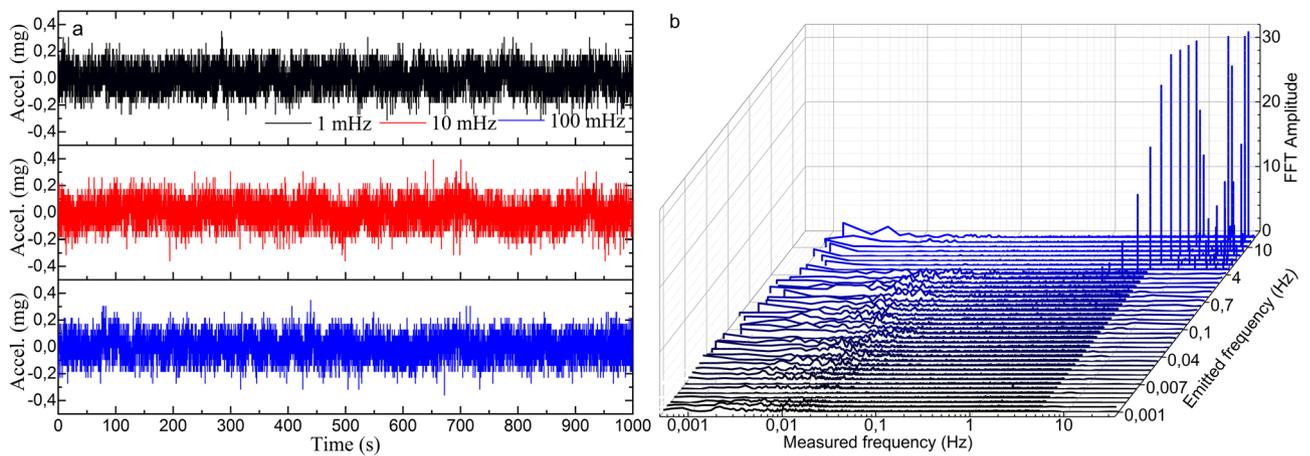

**Figure 8.** Results of the acceleration measurements in the electronic accelerometer. (**a**) Time response of three representative cases. (**b**) FFT amplitudes for frequencies from 30 Hz down to 1 mHz for a sinusoidal signal of 1 Vpp. The Measured frequency axis is in logarithmic scale.

Some other facts that should be highlighted from these results are the low variability and narrowness in amplitude and width, respectively, of the most prominent FFT component in all the cases (see Figs. 6b and 7b), with low level of the harmonic components. These characteristics are directly related to the purity of the acquired raw signal. This performance is critical for vibration measurements as it indicates that the device is practically insensitive to frequency variations if the same vibration amplitude is applied. This characteristic is noticeable if we pay attention to the time responses in Figs. 6a and 7a, where the recorded sinusoidal signal has practically the same amplitude in all the frequencies. According to the commercial accelerometer, it only detected vibrations of 2 Hz and above and with significantly noisier signal and with high level of harmonic components (see Fig. 8).

The second test consisted in emitting a sinusoidal signal of a fixed frequency (6 Hz) and varying its amplitude from 1 Vpp down to 10 mVpp (the lowest amplitude provided by the function generator) to define the LoD of each device in terms of amplitude of vibration, which is related to the acceleration of the oscillation movement. The time responses and FFT amplitudes of both devices are shown from Figs. 9 to 11. The optical device detected vibrations down to 10 mVpp above the established 3:1 SNR criteria. The noticeable progressive decrease in the amplitude of the signals in the time domain (see Figs. 9a and 10a) and the FFT (see Figs. 9b and 10b) is proportional to the diminishment of the amplitude of the emitted signal. In both cases, wavelength shift and power variations, the emitted signal can be clearly detected and the low level of the harmonic components is noticeable. Such results should be highlighted for the PD, whose FFT amplitudes are almost the double compared with the ones obtained by the spectrometer. In relation to the electronic accelerometer, according to the 3:1 SNR criteria, it detected the emitted signals from 1 Vpp down to 30 mVpp, which according to its calibration, covers an acceleration range from 29.64 to 0.76 mg. Its time response signals were significantly noisier (see Fig. 11a), and as a result of that, their corresponding FFT amplitudes were an order of magnitude below the ones of our MCF accelerometer (see Fig. 11b).

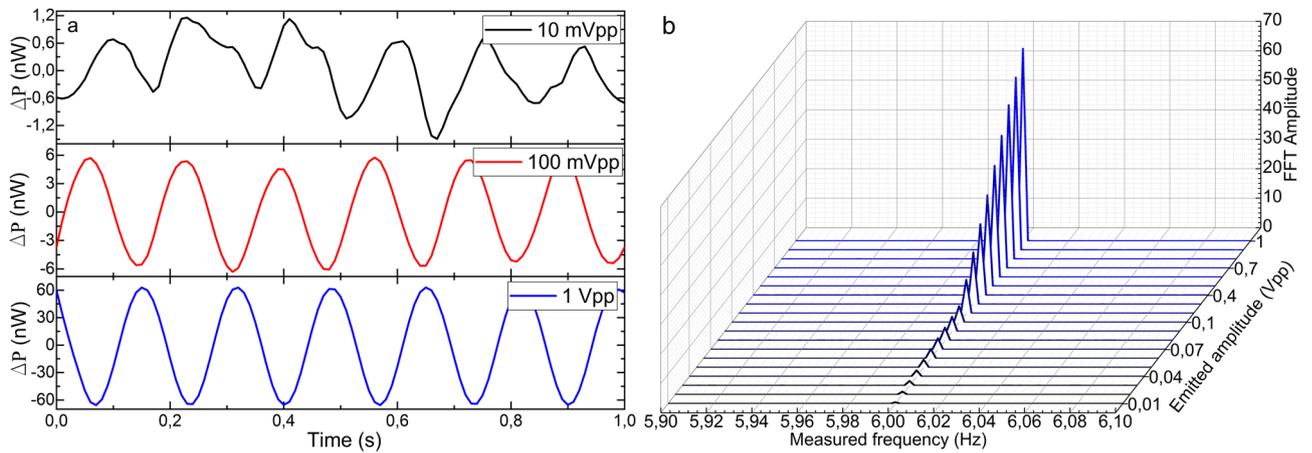

**Figure 9.** Results of the power variation measurements in the manufactured optical device. (**a**) Time response of three representative cases, and (**b**) FFT amplitudes for sinusoidal signals of 6 Hz and amplitudes from 1 Vpp down to 10 mVpp.

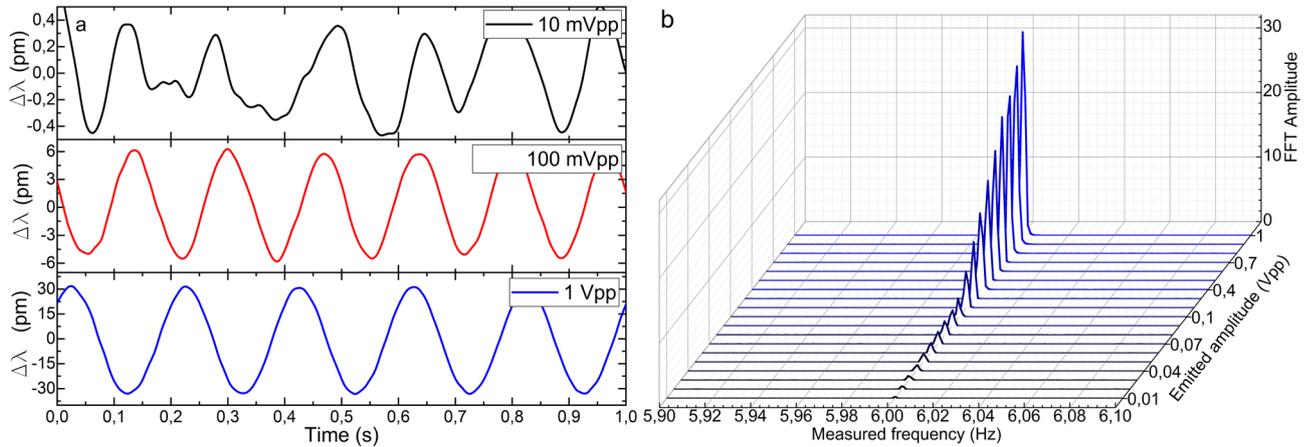

**Figure 10.** Results of the wavelength shift measurements in the manufactured optical device. (**a**) Time response of three representative cases, and (**b**) FFT amplitudes for sinusoidal signals of 6 Hz and amplitudes from 1 Vpp down to 10 mVpp.

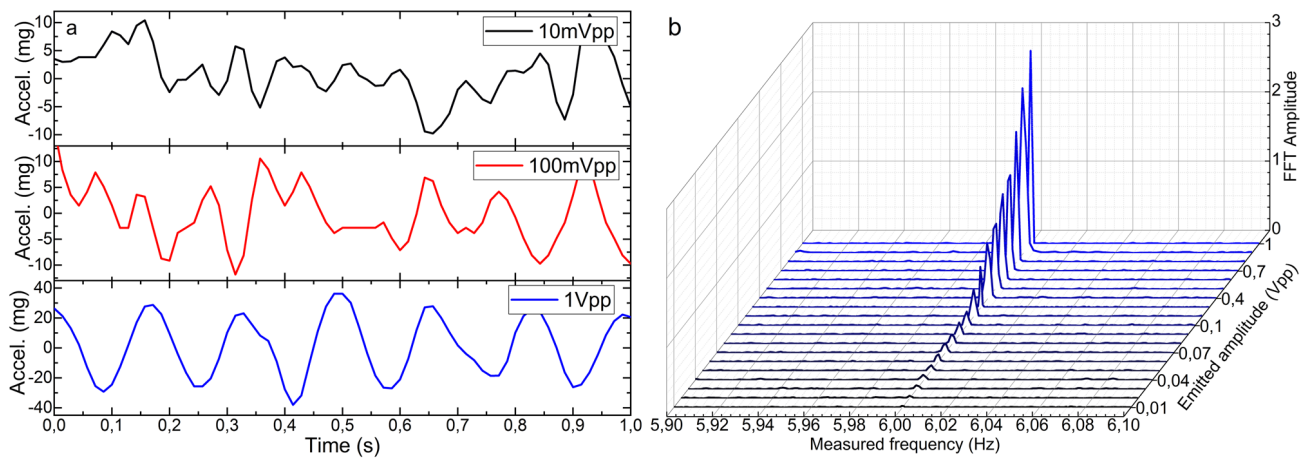

**Figure 11.** Results of the acceleration measurements in the electronic accelerometer. (**a**) Time response of three representative cases, and (**b**) FFT amplitudes for sinusoidal signals of 6 Hz and amplitudes from 1 Vpp down to 10 mVpp.

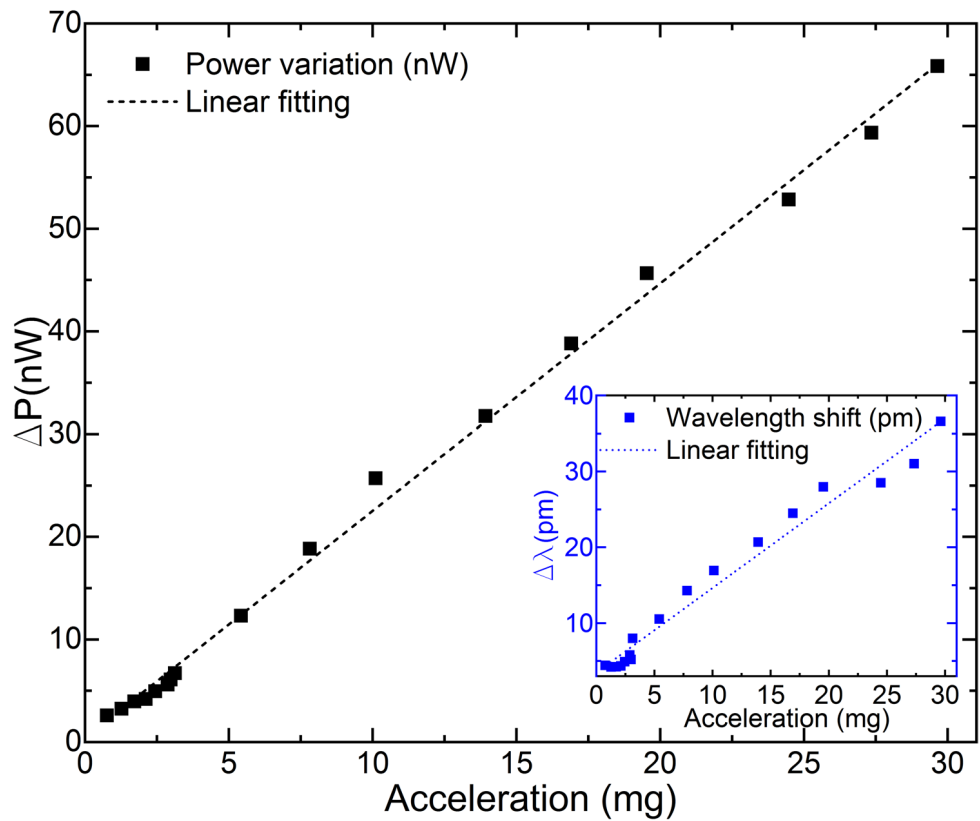

**Figure 12.** Calibration of the manufactured optical accelerometer in terms of power variation. The calibration of the device in terms of wavelength shift is shown in the inset.

The calibration resulting from these tests is shown in Fig. 12. The linear behavior of wavelength shift and power variations is significant, especially for the power variation measurements, where a sensitivity of 2.213 nW/mg with a Pearson squared correlation coefficient of $R^2 = 0.997$ and with a noise density of 1.083 µg/sqrt(Hz) was obtained. As a result, the correlation between the power variation (ΔP) and the acceleration (in mg) is as expressed in Eq. (6):

$$a = 0.450\Delta P - 0.143 \quad (6)$$

Equation 6 is applicable for accelerations from 29.64 mg down to 0.76 mg, as this was the LoD of the electronic accelerometer. However, considering that the MCF accelerometer detected vibrations of amplitudes below 30 mVpp and its significant linear behavior, we believe that this equation could be extrapolated and be valid for vibrations down to the tested limit (10 mVpp) and below. If so, this would indicate our device is capable of detecting accelerations of up to 0.25 mg. For real-field implementation, the robustness of the proposed measurement system could be improved by using a reference PD to monitor the stability of the light source in order to avoid any unwanted effect due to light power fluctuation.

According to wavelength shift measurements, a sensitivity of 1.116 pm/mg with a Pearson squared correlation coefficient of $R^2 = 0.976$ was achieved. It should be pointed out that our MCF accelerometer was optimized to operate with power variation measurements, which implied low sensitivity in terms of wavelength shift. Thus, such result points out that the device was surface bonded as close as possible as depicted in Fig. 5 and that it operates as intended.

## Conclusions

In this work, we have reported on a compact and highly sensitive all-fiber accelerometer based on two short segments of different lengths of asymmetric MCF. Such segments are rotated 180° with respect to each other and sandwiched between SMFs, creating a SMF-MCF1-MCF2-SMF structure. Its fabrication is fast, easily reproducible and customizable. Such configuration maximizes the change in the amplitude of the spectrum, which is related to power variation. Its interrogation is very simple and cost-effective, as it is made by few off-the-shelf equipment.

The manufactured device was subjected to vibrations of different amplitudes and frequencies, and its performance compared and calibrated with a commercial electronic accelerometer. It was found that our MCF accelerometer outperformed a commercial electronic accelerometer, as it was capable of detecting extremely low frequency vibrations down to 1 mHz with a sensitivity of 2.213 nW/mg, which makes it appealing for applications in which these characteristics are demanded, such as in seismology. To the authors´ best knowledge, this is the simplest optical fiber-based accelerometer that reaches this performance.

The MCF accelerometer proposed here is suitable for parallel multiplexing by means of an optical switch, which makes multi-point measurement feasible in order to cover large structures or areas. Thanks to the narrow reflection peaks provided by this SMF-MCF1-MCF2-SMF structure, several devices of this kind can be multiplexed in the same interrogation window. By modifying the length of the MCF segments in each device, the shape of the spectra and the location of the maxima can be customized individually, leading to an unambiguous identification of each. Moreover, the proposed structure may be embedded or surface bonded in oil pipelines or pillars, which facilitates its installation significantly, as it does not require expensive or complex setups. Lastly, we would like to highlight the potential of the device reported here to be direction sensitive by combining simultaneous analysis of wavelength shift and power variation. In this manner, the vibration as well as its direction could be identified accurately thanks to the effects observed in the spectrum.

Therefore, we believe that the MCF vibration sensor reported here may represent an alternative to conventional electronic and optical accelerometers thanks to its compactness, simplicity, high sensitivity, cost-effectiveness and versatility.

### Acknowledgments

Ministerio de Economía y Competitividad; Ministerio de Ciencia, Innovación y Universidades; European Regional Development Fund (PGC2018-101997-B-I00 and RTI2018-094669-B-C31); Gobierno Vasco/Eusko Jaurlaritza (IT933-16); ELKARTEK KK-2019/00101 (µ4Indust) and ELKARTEK KK-2019/00051 (SMART-RESNAK). The work of Angel Ortega-Gomez is funded by a PhD fellowship from the Spain Government. The work of Josu Amorebieta is funded by a PhD fellowship from the University of the Basque Country UPV/EHU.

### Author contributions

J.A. collaborated in the theoretical approach, performed the experiments, processed and analyzed data and wrote the first draft. A.O.-G. did the simulations and the theoretical approach, J.V. conceived and fabricated the device, designed and supervised the experiments. A.S., E.A.-L. and R.A.C. conceived and fabricated the MCF. R.F. developed the data acquisition software. G.D. and J.Z. supervised the experiments. All authors discussed the experimental data, revised and approved the manuscript. JA and JV wrote the final version with inputs of all the authors.

### Competing interests

The authors declare no competing interests.

### Additional information

**Correspondence** and requests for materials should be addressed to J.A. or J.V.

**Reprints and permissions information** is available at www.nature.com/reprints.

**Publisher's note** Springer Nature remains neutral with regard to jurisdictional claims in published maps and institutional affiliations.

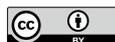